# The Case for a Hot Archean Climate and its Implications to the History of the Biosphere

David W. Schwartzman
Professor Emeritus, Department of Biology, Howard University, Washington D.C. 20059, U.S.
email: dschwartzman@gmail.com

**The case for a much warmer climate on the early Earth than now is presented. The oxygen isotope record in sedimentary chert and the compelling case for a near constant isotopic oxygen composition of seawater over geologic time support thermophilic surface temperatures prevailing in the Archean, with some support for hot conditions lasting until about 1.5 billion years ago, aside from lower temperatures including glacial episodes at 2.1-2.4 Ga and possibly an earlier one at 2.9 Ga. Other evidence includes the following:**
**1) Melting temperatures of proteins resurrected from sequences inferred from robust molecular phylogenies give paleotemperatures at emergence consistent with a very warm early climate.**
**2) High atmospheric $pCO_2$ levels in the Archean are consistent with high climatic temperatures near the triple point of primary iron minerals in banded iron formations, the formation of Mn-bicarbonate clusters leading to oxygenic photosynthesis and generally higher weathering intensities on land. These higher weathering intensities would not have occurred if seafloor weathering dominated the carbon sink, pulling down the temperature, hence this empirical evidence supports a hot climate and high carbon dioxide levels.**
**3) The inferred viscosity of seawater at 2.7 Ga is consistent with a hot Archean climate.**
**5) A cold Archean is hard to explain taking into account the higher outgassing rates of carbon dioxide, significantly smaller land areas and weaker biotic enhancement of weathering than present in the context of the long-term carbon cycle, taking into account the fainter Archean sun in climate modeling.**
**This evidence points to an important conclusion regarding biological evolution, namely to the critical role of a temperature constraint holding back the emergence of major organismal groups, starting with phototrophs, culminating with metazoans in the latest Precambrian.**

## Introduction

A long-standing debate in the Earth sciences has centered on the temperature history of the Earth's climate over geologic time. Was temperature then similar to the present or was it significantly higher, particularly for the Archean ? Critical in this debate is the interpretation of the empirical record of oxygen isotopes in sedimentary chert, silica deposited on the seafloor. The fractionation of oxygen isotopes (i.e., $O^{16}$ and $O^{18}$) between chert and seawater is a function of temperature at the time the isotopic record is established. If this fractionation is set during burial close to the temperature of the ocean near the seafloor then a climatic record can potentially be inferred if the seawater $O^{16} / O^{18}$ ratio is known along with knowledge of the ratio in chert.

**Discussion**

The hot climate interpretation of the oxygen isotopic record of marine Archean cherts (Knauth and Epstein, 1976; Knauth, 2005; Knauth and Lowe, 2003) has been commonly challenged by appeals to diagenetic resetting of the original isotopic ratios but this explanation is now considered implausible even by advocates of cooler climates (e.g., Kasting and Ono, 2006) in view of the robust case made by Knauth and Lowe (2003) and Knauth (2005). These Knauth papers also refute claims that the cherts used for support of a hot Archean climate are of hydrothermal origin (e.g., Paris et al., 1985).

A compelling case for a very warm Archean/early Proterozoic (50-70 deg C) (aside from the Huronian) can now be made. The empirical foundation for this temperature history is derived from the sedimentary chert oxygen isotopic record, specifically from the highest $O^{18}/O^{16}$ ratio at any age, since processes that would reset the climatic signature such as recrystallization with exchange with meteoric or higher temperature water during burial would decrease this ratio (Knauth, 2005). Alternatively, the shift in the range of $O^{18}/O^{16}$ ratios of cherts downward from the late Phanerozoic to the Archean by 10 per mil ($\delta O^{18}$ values; see Fig. 1) corresponds to a 45-50 deg C increase in climatic temperature or to the same shift in the seawater $O^{18}/O^{16}$ ratio at a constant temperature or to a mixture of both effects. Figure 1 shows this record with $\delta O^{18}$ values being a laboratory standard-corrected measure of the $O^{18}/O^{16}$ ratio. The upper bound corresponds to the ratio expressing the climatic signature.

The high paleotemperatures derived from the oxygen isotopic record of Archean cherts are contingent on the assumption of near constant isotopic oxygen composition of seawater over geologic time. If Archean seawater were sufficiently isotopically lighter than modern values the derived temperatures would be closer to a modern climate. A mechanism has been proposed for the necessary variation in the oxygen isotopic composition which requires shallower Archean ridgecrest depths than in the Phanerozoic, reducing high temperature seawater interaction with oceanic crust (Kasting and Ono, 2006; Kasting et al., 2006; Jaffres et al., 2007). However, the empirical record of ancient seawater-altered oceanic crust, including Archean eclogite proxies, appears to strongly support a near constant oxygen isotopic composition over geologic time.

In summary the evidence for near present seawater $O^{18}/O^{16}$ ratio back into the Archean includes the following:

1) Paleozoic seawater is not significantly lighter than the present: direct measurement of fossil seawater from salt inclusions (Knauth and Roberts, 1991). The same conclusion is reached from clumped isotope study on carbonate fossils (Came et al., 2007).

2) Inferences from the geologic record of ancient seawater-altered oceanic crust (ophiolites, their ecologite proxies, greenstones).

Buffering of oxygen isotopic ratio of seawater by reaction with oceanic crust over geologic time involves both low and high temperature reactions. At low temperatures the oceanic crust gains $O^{18}$ relative to seawater, while at high temperatures the oceanic crust loses $O^{18}$, coming closer to seawater $O^{18}/O^{16}$ ratio. This buffering has resulted in a near constant seawater $O^{18}/O^{16}$ ratio going back to the Archean (Muehlenbachs et al., 2003).

Table 1 shows data from Archean/Proterozoic ophiolites and the Archean Onverwacht greenstone which demonstrate the same oxygen isotope patterns as in the Phanerozoic, with depleted oxygen isotopic sections observed in the Onverwacht and Joruma. Subduction loss of the depleted sections is a likely explanation for the lack of preservation, e.g., in the Isua ophiolites. This subducted crust is represented in Archean ecologite proxies for depleted/enriched seawater-altered oceanic crust, with abundant depleted values, clear evidence of high temperature interaction with seawater (Jacob, 2004; Jacob et al., 2005). Additional support for this case is found in Schulze et al. (2003) and Spetsius et al. (2008). Type II ecologites preserve this signature, while the much more abundant Type I samples show the effects of metasomatism ((Greau et al., 2011. Further, altered mineral assemblages imply a higher temperature in the Archean upper oceanic crust than the modern (Nakamura and Kato, 2004).

**Table 1. $\delta\ O^{18}$ values of ophiolites, greenstones and ecologites**

$\delta\ O^{18}$ **Range**

| | | |
|---|---|---|
| **Phanerozoic ophiolites** | 3.2 - 13 | (Jacob (2004) |
| **Eclogite xenoliths from kimberlites** | 2 – 8 | (Jacob (2004) |
| **Isua ophiolite (3.8 Ga)** | 5.7 - 9.9 | (Furnes et al., 2007) |
| **Onverwacht greenstone (3.3-3.5 Ga)** | 3 -14 | (Hoffman et al., 1986) |
| **Joruma ophiolite (1.95 Ga)** | 1.4 - 8.5 | (Muehlenbachs et al., 2004) |

**Other evidence for near present seawater $O^{18}/O^{16}$ ratio back into the Archean/Hadean**

1) The seawater ratio inferred from the Hadean serpentinite (3.8 Ga) study (Pope et al., 2012).

2) $O^{18}$ is transferred to hydrosphere even by light meteoric water interaction in the case of Iceland, a subaerial, shallow ridgecrest (Hattori and Muehlenbachs, 1982; Gautason and Muehlenbachs, 1998), hence undermining the argument that postulated high ridgecrests would not permit loss of $O^{18}$ by the Archean oceanic crust.

3) Archean and Proterozoic volcanogenic massive sulfide deposits with ore fluids on the order of 300 deg C have measured $\delta\ O^{18} = 0$ per mil, again no evidence of very light seawater (personal communication, Muehlenbachs). Similarly, high temperature alteration phases in the ophiolites show no evidence of equilibration with $O^{18}$-depleted seawater (Turner et al., 2007). Hence the buffer was similar to now, $O^{18}/O^{16}$ seawater ratio close to present value.

4) As Robert and Chaussidon (2007) pointed out, Karhu and Epstein's (1986) chert-phosphate

pairs give δ O$^{18}$ seawater values close to present day (+/- 5 per mil) back into the Archean, even if the temperatures of equilibration for some samples may be higher than the sediment/seafloor interface (note: their cherts were not necessarily the heaviest for each age, hence diagenetic equilibration temperatures are probable for several samples).

5) Hydrogen isotopic composition inferred from δ D values measured from 2 Ga ophiolite-like complex (Lecuyer et al., 1996)

Though the oxygen isotopic record of carbonates is more easily reset than cherts, the upper bound to the δ O$^{18}$ values of carbonates over geologic time, assuming fractionation from the same seawater as cherts, agrees with the inferred climatic record derived from cherts (Figure 2). Therefore it is misleading to argue that the average carbonate δ O$^{18}$ values express a progressive trend towards more negative seawater δ O$^{18}$ going back into the Archean (e.g., Jaffres et al.., 2007).

**Silicon isotopes in cherts**

Silicon isotopes from marine cherts have been interpreted as supporting the hot climate scenario (Robert and Chaussidon, 2006). A commentary on the latter paper noted the correspondence of the upper temperature limit to photosynthesis to the estimated surface temperature derived from the chert oxygen and silicon isotopic record (De La Rocha, 2006). However, the interpretation of the silicon isotopic record of cherts has proved to be more complicated than argued by Robert and Chaussidon (2006), with numerous papers since this publication having discussed this isotopic record (van den Boorn et al., 2009; Heck et al., 2011; Marin et al., 2010; Marin-Carbonne et al., 2012, 2014; Geilert et al., 2014; Stefurak et al., 2015). Five criteria are proposed for robust reconstruction of seawater temperatures using micro-analysis of oxygen and silicon isotopes in chert samples by Marin-Carbonne et al. (2012). They conclude that only the 1.9 Ga Gunflint cherts, with an inferred seawater temperature of up to 52 deg C derived from oxygen isotopic analysis, satisfy these criteria from the limited selection of chert ages sampled.

**New evidence for a hot Archean climate from sedimentology**

The relative scarcity of ripple marks in thick sequences of 2.7 Ga deep-water turbidites imply a lower viscosity of seawater, with modeling indicating a seawater temperature range consistent with a hot Archean climate (Fralick and Carter, 2011).
.
**More recent challenges to the hot Archean climate**

1) Phase relations of iron minerals (magnetite, hematite, siderite) in Archean banded iron formations were analyzed by Rosing et al. (2010). They concluded that the Archean atmospheric carbon dioxide level was far too low to sustain a hot climate. Their assumption of simultaneous precipitation of these three iron minerals near the phase diagram triple point was challenged by Dauphas and Kasting (2011) and Reinhard and Planavsky (2011). Assuming for argument's sake that the three iron minerals were simultaneously precipitated, I submit their reasoning is circular, since they assumed a temperature near 25 deg C to derive their low carbon dioxide level. On the same phase diagram, hot Archean temperatures on the order of 60-80 deg C are consistent with much higher carbon dioxide levels in the atmosphere. The triple point at 70 deg C is above the faint early sun constraint: $pCO_2 = 0.3$ bars, log $pCO_2 = -0.52$. Further it should be noted that methanogens, inferred to be present in this environment by the authors, would likely reduce the $pCO_2$ level relative to the atmosphere.

**2)** Oxygen and hydrogen isotopes were analyzed in 3.42 Ga Archean cherts by Hren et al. (2009). They concluded that these cherts formed in waters below 40 deg C. However, the hydrogen extracted in their analysis was not likely derived from in-situ OH groups, rather is an artifact of sample preparation, and as such represented modern, not Archean hydrogen isotopic ratios. Pope et al. (2012) concluded their inferred Archean seawater δD values were not consistent with that of Hren et al. (2009)

3) The oxygen isotopes of Archean phosphate phases in chert were analyzed by Blake et al. (2010). They concluded that these phases crystallized in an ocean with similar temperature to the present (26 to 35 deg C). It should be noted that the analyzed phosphate sample with the highest $\delta O^{18}$ is described as consisting as predominately associated with iron oxides, "adsorbed, occluded and/or co-precipitated" with presumed particle size much smaller than the documented apatite inclusions (on the order of 10 microns). The oxygen isotopes of the host chert were not measured, which very likely equilibrated with the minute phosphate particles sampled during metamorphism, thereby changing the original oxygen isotopic ratio in these phosphate samples. Nevertheless, they concluded the Archean seawater oxygen isotopic composition was similar to the present, in direct contradiction to the significantly lower $O^{18}/O^{16}$ ratio inferred in Hren et al.'s (2009) analysis required to give their low temperatures.

4) Another very recent challenge is found in Som et al. (2012). They claim "Our result rules out very high Archaean ocean temperatures of 70 C–85 C …, because these would necessitate about 2–6 bar of carbon dioxide … plus 0.3–0.6 bar of water vapour, increasing barometric pressure far beyond the upper limit found here." They are referring to temperatures inferred from the earlier oxygen isotopic record of chert (3.5 to 3.0 Ga) assuming a $CO_2$-dominated greenhouse. They set an upper bound to Archean atmospheric density equal to 2.3 kg m$^{-3}$ based on modeling derived from presumed raindrops preserved in sedimentary rocks with an age of 2.7 Ga, not 3.0 to 3.5 Ga. However, Kavanagh and Goldblatt (2015) critiqued their methodology, setting an upper bound of 11.1 kg m$^{-3}$ or 4.8 times that of Som et al. (2012).

**Evidence for high Hadean/Archean $pCO_2$ levels in atmosphere/ocean**

Several recent papers have argued for high Hadean/Archean $pCO_2$ levels in the atmosphere/ocean in particular, Lichtenegger et al., 2010; Nutman et al., 2012; Dauphas et al., 2007; Rouchon and Orberger, 2008; Rouchon et al., 2010; Shibuya et al., 2010, 2012.

Dasgupta (2013) provides convincing evidence for these high levels:

"…owing to somewhat more efficient decarbonation of the subducting crusts at relatively shallow depths, the flux of $CO_2$ through Archean volcanic arcs likely was higher and deep subduction of carbon may have been hindered. In addition, owing to Earth's much hotter conditions, the subduction cycles may have also been episodic in nature for most part of the Precambrian …thus making subduction introduction of crustal carbon to the mantle even a less reliable ingassing process." "The comparison between the estimated temperatures of Precambrian subduction zones and the high-pressure experimental phase relations of crustal decarbonation suggests that significant $CO_2$ liberation at volcanic arcs likely helped to sustain higher $CO_2$ content in the exosphere for at least 2.5 billion years, between ~4 Ga and 1.5 Ga" "[Shibuya et al.'s 2012] study yielded an even more extreme flux of subducting carbon of ~$1.8 \cdot 10^{15}$ g of C/yr ($6.6 \cdot 10^{15}$ g of $CO_2$/yr). Again, all of this carbon being outgassed at volcanic arcs would yield at least 1-2 orders of magnitude higher flux than the total flux of modern volcanic $CO_2$ emission, with all magmatic centers combined. " (p. 207-208) "Massive release of $CO_2$ at Archean and Paleoproterozoic volcanic arcs thus may have supplied the necessary dose of greenhouse gas in the atmosphere to offset the dimmer early Sun and help sustain liquid water on Earth's surface." (p.219)

Further, high atmospheric/oceanic $pCO_2$ levels are apparent requirements for the formation of Mn-bicarbonate clusters leading to oxygenic photosynthesis at about 2.8 Ga, i.e., high bicarbonate and therefore high $CO_2$ levels are necessary (on order of 1 bar, pH of about 6.5) (Dismukes et al., 2001; Dismukes and Blankenship, 2005; Schwartzman et al., 2008).

The $pCO_2$ level required for 60 deg C climate at 2.7 Ga has been estimated at 1.6 bar of $CO_2$ for a total surface pressure of 2.74 bar; includes 1 bar $N_2$ (normal albedo) with 1 % $CH_4$ and $FH_2O = 5.6$ % (Personal communication, Feb. 5, 2013, Ramses Ramirez, Jim Kasting lab). This estimate was derived from a 1D climate model. A 3D GCM implies lower $pCO_2$ levels required for same temperature (Charnay et al., 2013), a conclusion supported by another 3D GCM study by Le Hir et al. (2014).

**But isn't the Archean $pCO_2$ level constrained to be < 0.14 bar by paleosol evidence as argued by Driese et al. (2011)?**

Following the methodology of Shelton (2006) and Sheldon and Tabor (2009), Driese et al. (2011) base their calculation of inferred atmospheric $pCO_2$ level on a formula with the formation time of the paleosol in the denominator (e.g., equation (13) in Sheldon, 2006). The critical problem in this approach is the assumption of a similar formation time for the Archean soil as the present, not taking into account the direct temperature effect on mineral dissolution as well as the evidence, both empirical and from modeling, that Archean weathering was much more intense than now. In other words, the conclusion that the $pCO_2$ level (and temperature) was only modestly higher than today was embedded in their assumptions.

Focusing on Driese et al.'s (2011) study on a 2.69 Ga paleosol, the derived limit is 10 - 50 ("best guess" = 41) times the present $pCO_2$ level. Sheldon (2006) assumes a present climatic temperature to calculate the rate of weathering in his mass balance approach. When a higher temperature (e.g., 60 deg C) at 2.7 Ga is used, the rate of weathering (inversely related to formation time) increases to 23-34 x the present (20 deg C), depending on the assumed temperature (60, 65 deg C; the temperature effect on weathering rate is exponential; assuming an activation energy of 70 kj/mole, and leaving out an additional factor, the runoff as a function of temperature; see discussion of temperature effect on chemical weathering of silicates in Schwartzman (1999, 2002), p.87-89, 150). In addition, another effect of higher temperatures would be to increase the $CO_2$ gradient and diffusion rate by more rapid reaction with CaMgFe silicates in zone C, the transition to fresh bedrock, thereby shortening the time to produce the paleosol.

In other words, a significantly higher assumed temperature at 2.7 Ga would have increased the weathering rate relative to today's climate, and significantly reduced the formation time necessary to create the observed paleosol. Assuming a climatic temperature of 60-65 deg C the computed atmospheric $pCO_2$ is on the order of 1000 PAL, the product of Sheldon's high limit of 50 x the temperature enhancement factor, with other inputs unchanged, noting that the variation with temperature of the Henry's Law and diffusion constant for $CO_2$ is much smaller. Hence, the conclusion (low $pCO_2$, low temperature) appears to be embedded in the assumptions used to compute the result.

In an experimental study aimed at simulating high temperature and $pCO_2$ weathering, Alfimova et al. (2014) concluded that the atmospheric $pCO_2$ level was "not much higher than 25 PAL", based on a comparison of reaction-produced and paleosol mineralogies. However, no experiments were made combining high temperature (50 and 75 deg C) and high $pCO_2$ (1 and 10 bars). Further, the only Archean paleosol used in this comparison was the Mt. Roe (2.76 Ga), with the rest in the 2.0 – 2.45 Ga age range, corresponding to the Huronian climatic interval. Moreover, the Mt. Roe paleosol has been metasomatized with addition and subtraction of several elements from the presumed original composition weathered from basalt, with both Rb-Sr and Sm-Nd ages reset to the Huronian age interval (MacFarlane et al., 1994)

**More challenges and evidence for a very warm Archean**

**1) The Archean was not hot given the Pongola glaciation at 2.9 Ga**

The glacial origin of the Pongola diamictites (2.9 Ga) is supported in Young et al. (1998), but are there alternative explanations for this deposit? Here are some alternatives:

a) Given the mounting evidence for a vigorous impact history in the Archean (Glikson, 2010) and the arguably weaker case for the glaciation at 2.9 Ga in contrast to the Huronian, a potential alternative explanation is that the Pongola diamictites are reworked impact debris, including inherited striations normally attributed to glaciation. Striations can be produced by impacts (Rampino et al., 1997). Debris flows can have structures looking like dropstones without being of glacial origin (Kim et al., 1995).

b) Pongola diamictites are hypercane deposits; hypercanes were postulated to occur over the ocean at temperatures at 50 deg C or above with speeds are postulated to reach over 800 km/h. (Emanuel et al., 1995; Emanuel, 2003). If confirmed, striations on clasts thought to be the result of cold temperatures are actually the result of hot temperatures. This scenario was originally suggested to me by Euan Nisbet in 2000.

Or alternatively, could glaciation at 2.9 Ga be an excursion from a hot Archean climate before and after ? Two potential scenarios are suggested:

a) Oblique impact creation of an orbiting ring around Earth, lasting a few million years, leading to dramatic cooling (Schultz and Gault, 1990), perhaps complemented by reflective $CO_2$-ice cloud formation.
b) A transient but severe reduction in volcanic outgassing of $CO_2$ generates a much colder climate as a steady-state in the long term carbon cycle (analogous to what Condie et al., 2009 proposed for the beginning of the Proterozoic at 2.45 Ga). Such a transient would have to occur earlier than 2.9 Ga with sufficient time to sequester 1-2 bars of $CO_2$ from the atmosphere/ocean into the crust. Perhaps impact generation of massive regolith deposits, increasing crustal weatherability, could have facilitated this drawdown of $CO_2$ even without a large reduction of volcanic outgassing.

**2) Generally higher weathering intensities**
Evidence includes paucity of Archean arkoses (sandstones with feldspars), and a higher chemical weathering intensity index ("CWI") (See discussion in Schwartzman et al., 2008, citing Corcoran and Mueller, 2004; Condie et al., 2001; also Simpson et al., 2012).  An estimate of CWI can be inferred from the following relationship:
CWI = $(V/V_o)$ x $(A_o/A)$, where V is volcanic outgassing flux of $CO_2$, $V_o$ now, and A is exposed land area available for weathering, $A_o$ now.  For plausible Archean values of $V/V_o$ = 2 and $A_o/A$ = 10 (especially for $\geq$2.7 Ga), CWI = 20 times the present chemical weathering intensity.  Note that the chemical weathering sink flux balancing the volcanic source flux of $CO_2$ is 2 times the present, with V = chemical sink flux. In other words, much smaller exposed continents in the Archean must do the work of balancing a higher volcanic outgassing flux, with much more modest biotic enhancement of weathering (BEW) than in more recent time, hence much higher weathering intensities. The inferred BEW for the present and its progressive increase to this value over geologic time emerges as a model result with reasonable assumptions regarding continental growth and outgassing rates (Schwartzman, 1999, 2002; see Figure 3); the Archean (and Hadean) was near abiotic with respect to BEW, so this factor must be taken into account in modeling the long-term carbon cycle.

**3) A cold Archean is hard to explain in the context of the long term carbon cycle**

Noting that with higher outgassing rates of carbon dioxide, smaller land areas and weaker biotic enhancement of weathering in Archean than present, taking into account the fainter Archean sun in climate modeling and higher weatherability of mafic crust. a cold Archean is hard to explain (Schwartzman, 1999 2002). The generally higher Archean weathering intensities imply seafloor weathering (e.g., Sleep and Zahnle, 2001) was **not** driving the C sink as thereby reducing temperature, hence a hot climate and high carbon dioxide/methane levels.

Why is there now the equivalent of 50-60 bars of $CO_2$ in the crust as carbonate and reduced organic carbon, if seafloor weathering/carbonation of oceanic basalt, especially in the early Precambrian, drew down the atmospheric $CO_2$, sinking most of it in the mantle, instead of degassing it back to the atmosphere/ocean during subduction?

Modeling the history of juvenile degassing can explain how this 50-60 bar equivalent of crustal sedimentary carbon is consistent with high steady-state atmospheric $pCO_2$ levels and surface temperatures prior to about 2.5 Ga. (see p.364-5, Schwartzman and Volk, 1991).

**3) The Huronian is followed by no apparent glaciations for over a billion years (2 Ga to NeoProterozoic snowball)**

Actual isotopic evidence for glacial temperatures in the Huronian is apparently derived from oxygen isotopic analysis of Karelian metamorphic rocks, with very depleted $O^{18}$ signatures (Bindeman et al., 2010; Bindeman and Serebryakov, 2011); Vysotskiy et al., 2014) although this glacial age interpretation has been challenged with the argument that the signature corresponds to the protolith at the time of 2.9 Ga Pongola glaciation (Krylov, 2008; Krylov et al., 2011). The Huronian glaciations are plausibly explained by decline of volcanism at 2.45 Ga (Condie et al., 2009; Eriksson and Condie, 2014) and rise of emerged continental area. Condie's papers argue that volcanism dramatically declined between 2.45 and 2.2 Ga in a "magmatic shutdown. A simple model of the long term carbon cycle implies that a significant decline in volcanic outgassing of $CO_2$, starting at 2.45 Ga, combined with exposed land area less than 20% the present owing to flooded continents prior to the Proterozoic (Flament et al., 2013) is consistent with cooling to a new steady-state at modest temperatures. Roughly 50 million years before the rise of atmospheric oxygen a decline of volcanism could generate a steady-state with a lower atmospheric $CO_2$ level and modest climatic temperature so that the rise of atmospheric oxygen could wipe out most of the greenhouse methane and thereby drive down the temperature further to bring on the Huronian glacial episodes. A hot climate returned as volcanic outgassing rates increased around 2 Ga (note the isotopic evidence for a 50 deg C climate (Winter and Knauth, 1992; Marin et al., 2010). Teitler et al. (2014) have proposed a very similar explanation for the Huronian glaciations.

**4) Highly stratified oceans apparently persisting until the mid Proterozoic, with relatively hot climates being responsible** (see Lowe, 1994)

**5) Evidence from Molecular Phylogeny**

Note the apparent absence on molecular phylogenetic trees of deeply-rooted mesophiles/ psychrophiles shown in Figure 4. If Archean temperatures were similar to the Phanerozoic, then some of the low-temperature prokaryotes should be grouped near the root with the hyperthermophiles/thermophiles (Schwartzman and Lineweaver, 2004a,b).

The inferred paleotemperatures from resurrected (elongation) proteins and their melting temperatures (Gaucher et al., 2008) are consistent with hot Archean, e.g., cyanobacteria emerging at about 60 deg C at 2.8 Ga, as also proposed by Schwartzman et al., 2008). More support for hot

Archean temperatures can be found in other recent molecular studies (Boussau et al., 2008; Perez-Jimenez et al., 2011).

## 6) Sequence of organismal emergence times and their upper temperature limit for growth (Tmax)

Tmax values at emergence are consistent with the chert paleotemperatures, and the more recent emergence of psychrophiles, hence little or no ice before the step down in temperature at about 1.5 Ga (aside from the Huronian; mesophiles emerging then would be likely wiped out by a return to near thermophilic temperatures) (Table 2).

**Table 2. Upper temperature limits for growth of living organisms, approximate times of their emergence**

| Group | Approximate upper temperature limit ("Tmax") (deg C) | Time of Emergence (Ga) |
|---|---|---|
| "Higher" kingdoms: | | |
| Plants | 50 | 0.5-1.5 |
| Metazoa (Animals) | 50 | 0.6-1.5 |
| Fungi | 60 | 0.6-1.5 |
| Eucaryotes | 60 | 2.1-2.8 |
| Procaryotes | | |
|    Phototrophs | 70 | ≥3.5 |
|    Hyperthermophiles | >100 | >3.8 |

(Temperatures from Brock & Madigan, 1991)

## 7) Oxygen levels alone cannot explain the big delay in the appearance of the "Higher" Kingdoms, in particular Metazoa

I proposed a temperature constraint for metazoan emergence in Schwartzman (1999, 2002). Mills and Canfield (2014) have also questioned whether the oxygen level delayed the first emergence of metazoa. It should be noted that the presence of high oxygen in surface microenvironments such as cyanobacterial mats, likely went back to 2.7-2.8 Ga. Further, the role of a temperature constraint for metazoan emergence is supported by the recent discovery of living anoxic metazoa (Danovaro, R. et al., 2010) and evidence for anaerobic metabolism in the common ancestor of eucaryotes (Muller et al., 2012). Given a Tmax for eucaryotes near 60 deg C, this conclusion implies an Archean emergence time. An atmospheric $pO_2$ constraint is indicated for the emergence of macroeucaryotes, including metazoa in the Phanerozoic (Berner et al., 2007).

**Some suggestions for future research**

Levin et al. (2014) proposes an approach to resolving the long-standing debate regarding the hot Archean climatic interpretation of the oxygen isotopic record of cherts utilizing triple oxygen isotope analysis. Micro-analysis of oxygen and silicon isotopes in cherts is now a common research method (e.g., Heck et al., 2011; Marin et al., 2010; Marin-Carbonne et al., 2012, 2014; Geilert et al., 2014; Stefurak et al., 2015; Cunningham et al., 2012). The recent measurement of the isotopic composition of paleoatmospheric argon, in particular the $Ar^{40}/Ar^{36}$ ratio of the Archean atmosphere (Pujol et al., 2013) opens up a potential research direction, namely combining oxygen, silicon and Ar40/Ar39 isotopic analysis on the same micro-sample. The identification of a paleoatmospheric Ar signature would strongly support preservation of near equilibrium with Archean seawater in the silica analyzed. This micro-approach could potentially avoid the already documented problem of radiogenic $Ar^{40}$, in-situ produced from $K^{40}$ decay and inherited from hydrothermal sources, found in bulk chert samples that elevate the initial $Ar^{40}/Ar^{36}$ ratio above the ancient atmospheric value (e.g., Alexander, 1975; Sano et al., 1994).

**Conclusions**

A major implication of this climatic temperature history is that temperature itself is a constraint on the emergence of major organismal groups. As a result of the co-evolution of life and its abiotic environment, I have argued that the evolution of Earth's biosphere is close to being deterministic, i.e., its origin and history and the general pattern of biotic evolution are very probable, given the same initial conditions, potentially a model for Earth-like planets around Sun-like stars (Schwartzman, 1999, 2002).

**Acknowledgments**

Paul Knauth is responsible for the foundational research underlying this paper. He provided invaluable insights regarding the interpretations of the Hren et al. (2009) and Blake et al. (2010) papers. I thank the participants for their insights given at the Faint Early Sun Conference (Space Telescope Science Institute), April 2012, Baltimore, Maryland, where I presented an earlier argument for a hot early Earth climate.

**Figure Legends**

**Figure 1. Oxygen isotope variation in cherts with Age.** Data from Knauth, 2005.

**Figure 2. Oxygen isotope variation in sedimentary carbonates with Age.**
(from L. Paul Knauth)

**Figure 3. Temperature History of the Biosphere.** A first order temperature curve is shown, with a likely uncertainty of about +/- 10 deg C. Emergence times of organismal groups are shown. The numbers on the curve are the model-derived ratios of the present biotic enhancement of weathering (BEW) to BEW values at indicated times (Schwartzman, 1999, 2002). Trends are from 4.0 Ga to now.

**Figure 4. Phylogenetic tree based on rRNA sequences, color coded to the maximum temperature of growth (Tmax).** Lineweaver and Schwartzman (2004) with phylogeny based on Pace (1997).

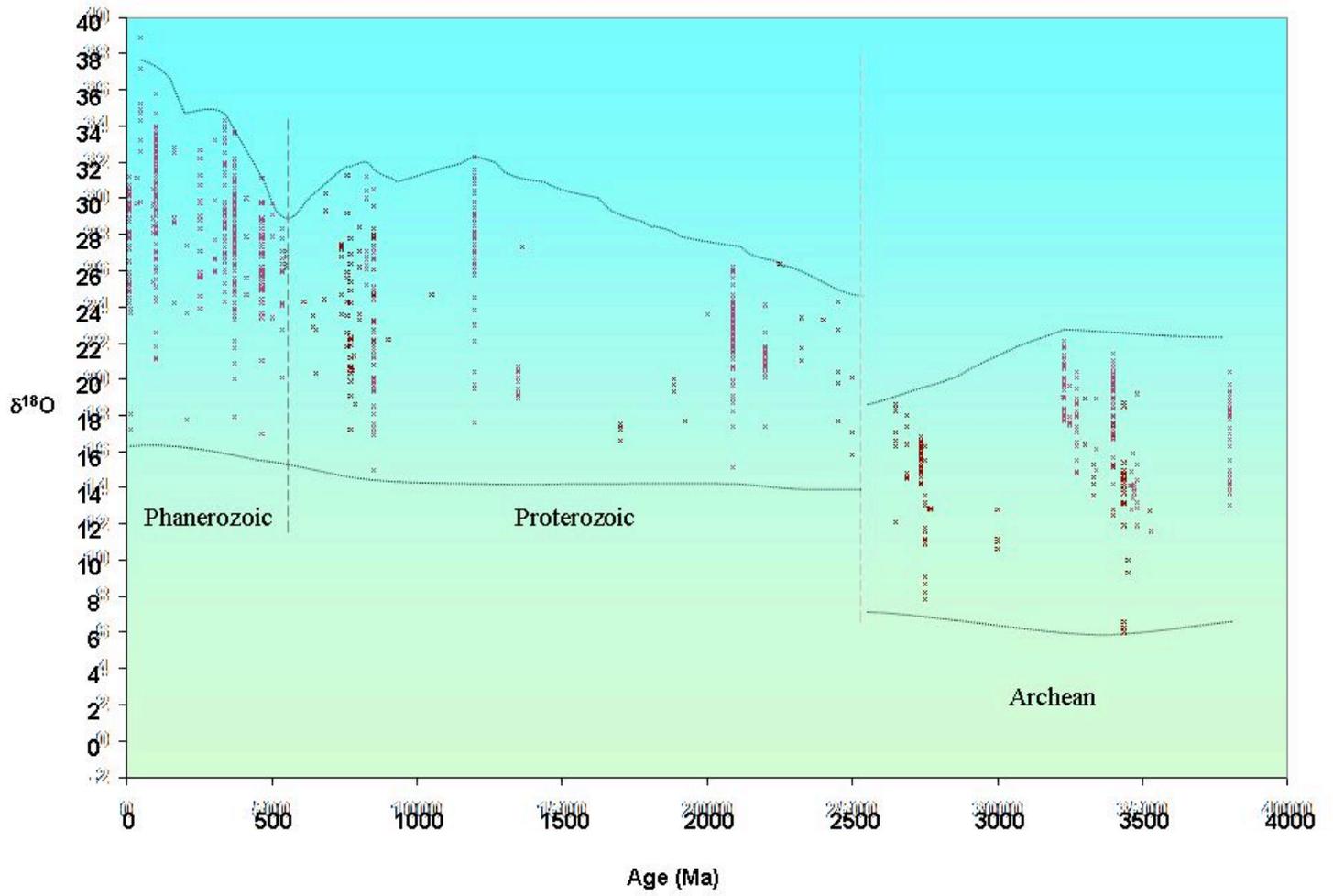

**Figure 1**

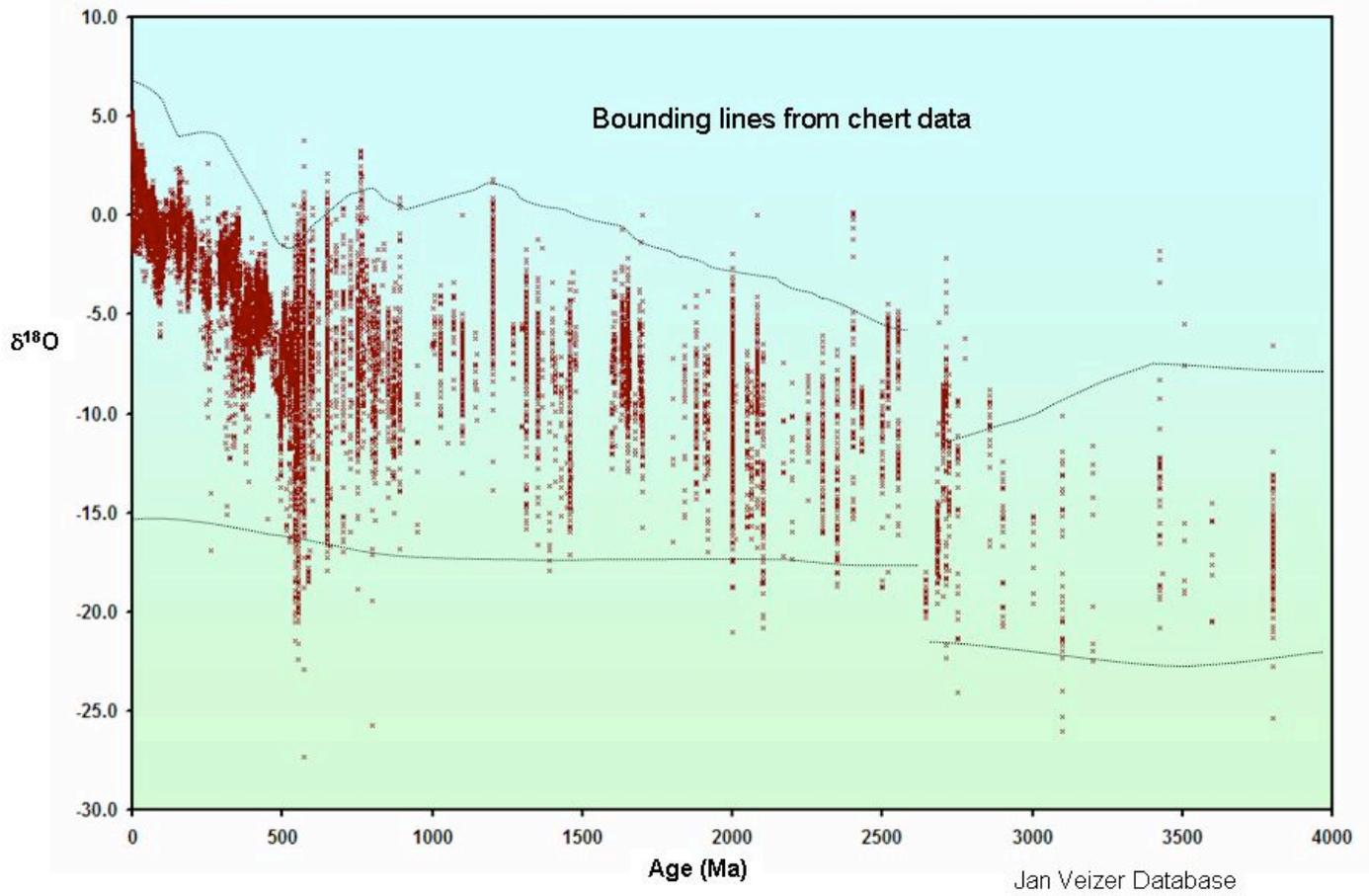

**Figure 2**

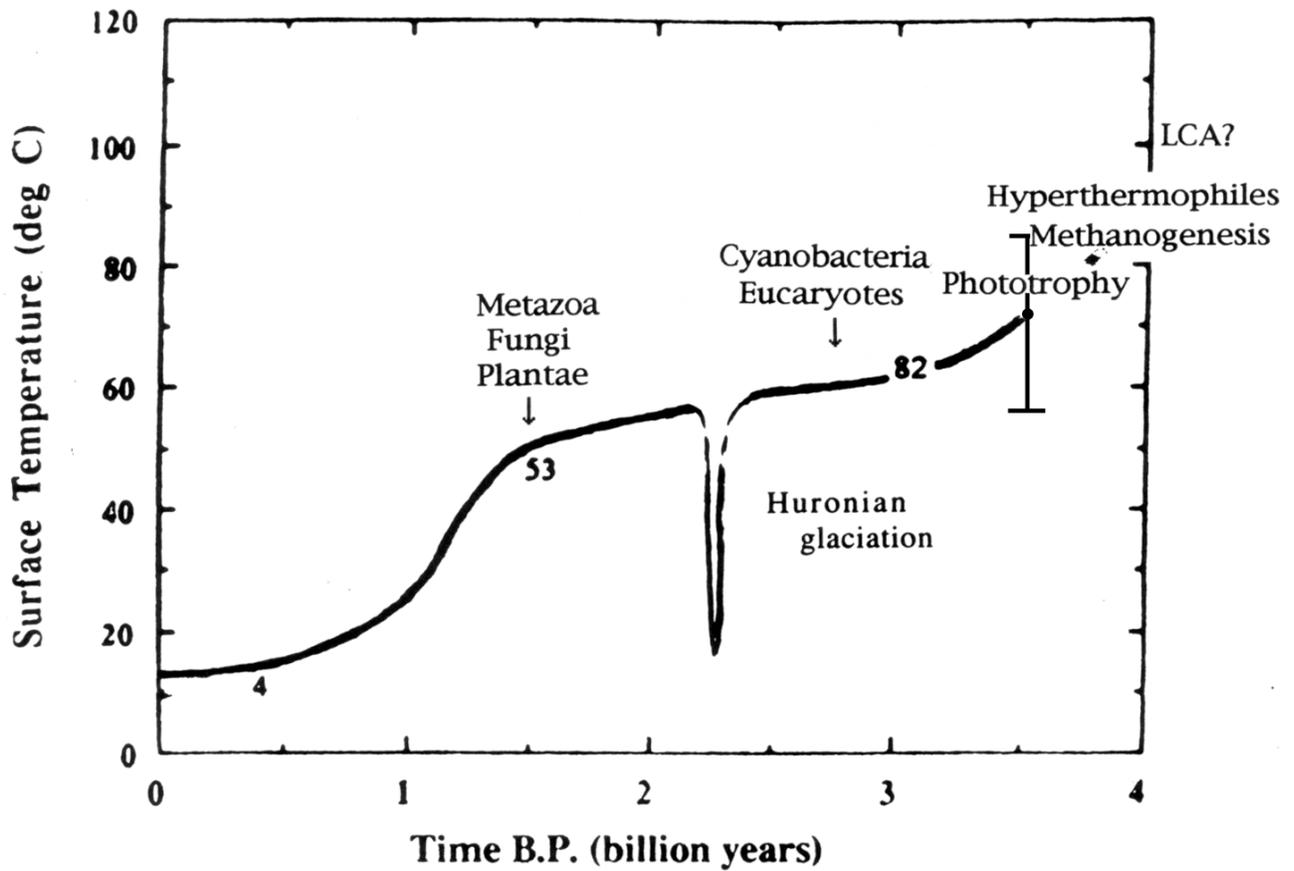

**Figure 3**

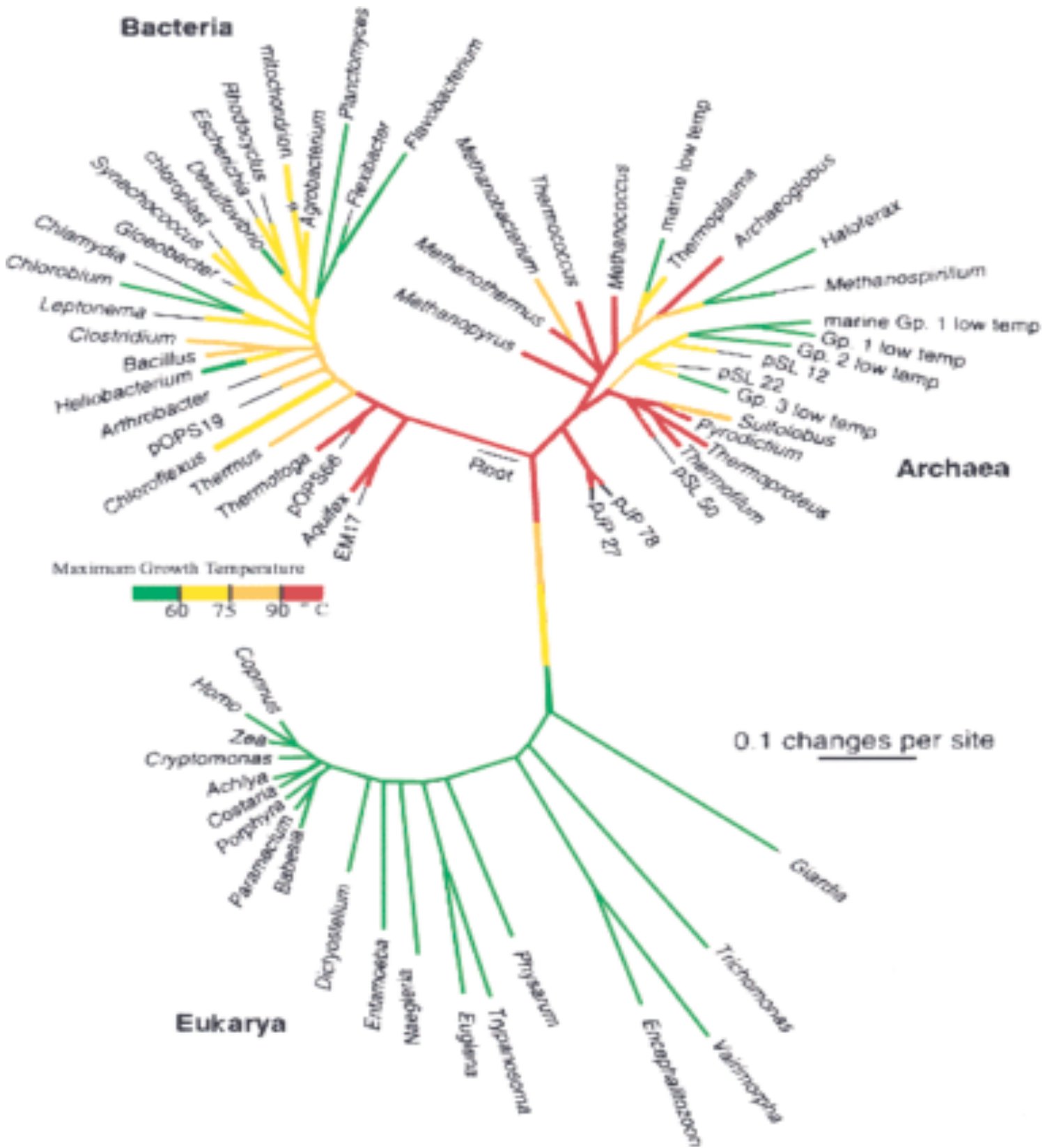

**Figure 4**